\documentstyle[12pt]{article}

     \newcommand{\eq}{\begin{equation}}

     \newcommand{\en}{\end{equation}}

     \newcommand{\spa}{\hspace{0.5cm}}
     \newfont{\myfont}{msbm10 scaled\magstep1}
     
\begin{document}

      \begin{titlepage}

     \rightline{PUPT-1331}
     \rightline{July 1992}

     \vskip 10mm

 \centerline{\bf \LARGE
 The New Charged Massless States }
 \vskip 3pt
\centerline{\bf \LARGE   of Quantum Electrodynamics }

    \vskip 2.0cm

    \centerline{\sc E.C.Marino \footnote{On sabbatical leave from
Departamento de F\'{\i}sica, Pontif\'{\i}cia Universidade
Cat\'{o}lica, Rio de Janeiro, Brazil. E-mail :
marino@puhep1.princeton.edu }}

     \vskip 0.6cm

     \centerline{\it Joseph Henry Laboratories }
    \centerline{\it   Princeton University }
    \centerline{\it    Princeton , NJ 08544 }

     \vskip 2.5cm

  \begin{abstract}  Quantum Electrodynamics can be formulated
as the theory of an antisymmetric tensor gauge field. In this
formulation the topological current of this field appears as an
additional source for the electromagnetic field.The topological
charge therefore acts physically as an electric charge. The
topologically nontrivial, electrically charged, sector contains massless
states orthogonal to the vacuum which are created by a gauge invariant
local operator and can be interpreted as coherent states of photons.
\end{abstract}

\vfill
\end{titlepage}

\leftmargin 25mm

\topmargin -8mm
\hsize 153mm

\baselineskip 7mm
\setcounter{page}{2}
\spa The  use of nonperturbative methods in quantum field theory
has always revealed surprising and unsuspected features. Despite
the tremendous success of perturbation theory in describing the
physical properties of a variety of systems, including
the gauge theories of the fundamental interactions
, it is well known that a wealth of new
phenomena can be uncovered by nonperturbative
investigations.
A very interesting class of nonperturbative problems in quantum
field theory is the one concerning the existence of topological sectors
in the Hilbert space. These sectors are characterized by the fact that
the states they contain bear a charge whose associated current is
identically conserved. In the most frequent situations there are
classical finite energy solutions of the field equations (solitons) carrying
the topological charge and manifesting themselves as massive quantum
states in the topological sector. In less common cases, despite the
absence of
classical finite energy solitons the theory still has quantum states
(usually massless) carrying the topological charge.

\spa In the present work, we study Quantum Electrodynamics in 3+1 D. We
show
 that the vector field (photon) part of the
 theory possesses a hidden topological sector, whose
topological charge behaves as the electric charge itself. In spite
of the fact that the field equations do not admit classical
finite energy solutions, we construct an
operator which creates the states in this hidden sector and
 evaluate
its correlation functions. From the large distance behavior of
these correlation functions we can deduce that the topological  charge
carrying states
are massless and orthogonal to the vacuum.

\spa There is actually an extremely simple system which displays
essentially the same features we discovered in QED,
namely the massless scalar field in two-dimensional spacetime. The
field equation $\Box \phi =0$ implies the conservation of the
current $j^{\mu}=\partial^{\mu} \phi$. It also
implies the identity $j^{\mu}=\tilde{j}^{\mu}$, where $\tilde{j}^\mu =
\epsilon^{\mu \nu } \partial_\nu \tilde{\phi}$ is the topological current
of the scalar field $\tilde{\phi}$ which is itself defined by the
above identity between currents. It is clear that no
polynomial in $\phi$ could create states carrying the charge associated
with the current $j^\mu$. It is not difficult
to see, however, that the
operator $\exp\{ib \phi \}=\exp\{ib\int ^x \dot{\tilde{\phi}}dx\}$
 carries b units of this charge and creates (massless)
states orthogonal to the vacuum even though the
theory possesses no classical solitons. We will see that the photon
sector of QED presents a structure completely analog to the one
just described.

\spa Let us start showing that QED
 can be formulated in terms of an antisymmetric tensor
gauge field.
 Indeed, starting from
$$
  {\cal L}_{W}+{\cal L}_{I}
 = -(1/12) W_{\mu \nu \alpha}(-\Box)^{-1}W^{\mu \nu
\alpha} -(1/2) \epsilon^{\mu \nu \alpha \beta}A_{\mu}\partial
_{\nu}W_{\alpha \beta}
 $$
where $ W_{\mu \nu \alpha} = \partial _{\mu}W_{\nu \alpha}+
\partial _{\nu}W_{\alpha \mu}+\partial _{\alpha}W_{\mu \nu}$ is the
field intensity tensor of the antisymmetric field,
we can readily obtain the pure Maxwell lagrangian
$
    {\cal L}_M = -(1/4) F_{\mu \nu}F^{\mu \nu}
$
upon functional
integration over $W_{\mu \nu}$.
We will therefore write the photon part of the QED lagrangian as
   $ {\cal L} =(1/2) [ {\cal L}_{W}+{\cal L}_{I}+{\cal L}_M]$. Integration
over $W_{\mu \nu} (A_{\mu})$ in ${\cal L} $
will produce the lagrangian ${\cal L}_M ({\cal L}_{W})$ (for the gauge
fixing terms, see below). The nonlocality of ${\cal L}_{W}$ is only
apparent since $(-\Box )^{-1}$ has support on the light-cone surface
 (we choose the Feynman prescription).
The field equations associated with this lagrangian are
\eq
     \partial _{\nu}  F^{\nu \mu} = (1/2) \epsilon ^
{\mu \nu \alpha \beta}\partial _{\nu}W_{\alpha \beta}\ \ \ \ \ \ \ \
\partial _{\alpha}W^{\alpha \mu \nu} = (-\Box)\epsilon ^
{\mu \nu \alpha \beta}\partial _{\alpha}A_{\beta}
\label{ecampo1}
\en
We immediately see that the topological current $J^{\mu}\equiv
\frac{1}{2}\epsilon ^{\mu \nu \alpha \beta}\partial _{\nu}W_{\alpha
\beta}$ becomes a source for the electromagnetic field.
We could as well consider the coupling of $A_{\mu}$
to a matter current $j^{\mu}_{m}$. In this case we would have the sum
$J^{\mu} + j^{\mu}_{m}$ as the total source of the electromagnetic field
in the Maxwell equation in (~\ref{ecampo1}).
The analogy with the 1+1 D massless scalar field is established
by: $A_{\mu}\leftrightarrow \phi$,\ $W_{\mu \nu }
\leftrightarrow \tilde{\phi}$,\  $\partial _{\nu}F^{\nu \mu}
 \leftrightarrow
  \partial ^{\mu}\phi$,\ $J^{\mu}\leftrightarrow \epsilon ^{\mu \nu}
\partial _{\nu}\tilde{\phi}$.

\spa It is not difficult to find a classical
configuration of the antisymmetric field having nonzero topological
charge. Consider, for instance the configuration \cite{gamb}:\
$
\overline{W}_{ij} = -\frac{1}{4\pi}\epsilon _{ijk}\frac{{\rm x}^{k}}{|\vec{\rm
x}|^{3}}
,\ \ \  \overline{W}_{0i} = 0
$.
For this we have $J^{0} = \delta ^{3}(\vec{\rm
x})$ and $J^{i} = 0$ and therefore unit topological charge. The energy
of this configuration however is infinite. This would be exactly the
situation if we tried to probe a Sine-Gordon soliton in our 1+1 D
theory of the massless scalar field $\tilde{\phi}$
: it would have nonzero topological
charge but infinite energy.

\spa The field equations (~\ref{ecampo1})  are
solved by
\eq
F^{\nu \mu}=(1/2)\epsilon ^{\mu \nu \alpha \beta}W_{\alpha \beta}
+\Lambda ^{\mu \nu} \ \ \ \ \
W^{\alpha \mu \nu}=(-\Box)\epsilon ^{\mu \nu \alpha \beta}A_{\beta} +
\Lambda ^{\alpha \mu \nu}
\label{sol1}
\en
where
$\Lambda ^{\mu \nu}=-(1/2)\epsilon ^{\mu \nu \alpha \beta}(\partial
_{\alpha}\Lambda _{\beta}-\partial _{\beta}\Lambda _{\alpha})$ and
$\Lambda ^{\alpha \mu \nu }=-(-\Box)\epsilon ^{\mu \nu \alpha \beta}
\partial _{\beta}\Lambda$ for arbitrary $\Lambda _{\beta}$ and $\Lambda$.
We see that we can always choose these last two quantities so as to
cancel the variations of
 $W_{\alpha \beta}$ and
$A_{\beta}$, respectively, under gauge transformations
 in such a way that the gauge invariance of the
l.h.s. of (~\ref{sol1}) is guaranteed. Inverting the first equation in
(~\ref{sol1}) and choosing the Lorentz gauge for both the
electromagnetic and antisymmetric fields, namely, $\partial _{\mu}
A^{\mu}=0$ and $\partial _{\mu}W^{\mu \nu}=0$ we have
$$W^{\mu \nu}=(1/2)\epsilon ^{\mu \nu \alpha \beta}F_{\alpha
\beta}\ \ \ \   W^{\alpha \mu \nu }=(-\Box)\epsilon ^{\mu \nu
\alpha \beta}A_{\beta}$$
(Note that since in the Lorentz gauge $\Box\Lambda=0$, $W^{\alpha \mu
\nu}$ remains invariant under $A_{\mu}\rightarrow A_{\mu}+\partial
_{\mu}\Lambda$).
We see that in this particular gauge the antisymmetric
field can be identified with the dual field intensity tensor
$\tilde{F}^{\mu \nu}$ and the Lorentz gauge condition (on
$W^{\mu \nu}$) becomes the
Bianchi identity (observe that the configuration $\overline{W}_{\mu
\nu}$
corresponds to the field of a static point electric charge).

\spa We can now obtain the basic commutation rules that will be needed
later. Using a covariant gauge (Lorentz) quantization (see \cite{ty} ,
for instance) we have the equal-time commutation relations
 $
[A^{i}(x),E_{j}(y)]=i\delta ^{i}\ _{j}\delta ^{3}(x-y)
$
where $E^{i}=F^{i0}$ is the electric field. Using this equation and
 the Lorentz gauge solution for W in terms of $A_{\mu}$ we find
$$
[W^{0ij}(x),W_{kl}(y)]=i\Delta ^{ij}\ _{kl}(-\Box)\delta ^{3}(x-y)
\ \ ;\ \
[W^{0ij}(x),E^{k}(y)]=i\epsilon ^{ijk}(-\Box)\delta ^{3}(x-y)
$$
where $\Delta ^{ij}\ _{kl}\equiv\delta ^{i}\ _{k}\delta ^{j}\ _{l}-
\delta ^{i}\ _{l}\delta ^{j}\ _{k}$.

\spa Let us introduce now the operator $-\ \mu\ \
 -$ which will create the states
carrying topological charge. This belongs to a class of operators which
was introduced in two, three, and four
dimensional spacetime as the creation operators of the respective
topological excitations, namely$-$ kinks, vortices and magnetic monopoles
\cite{kinks,emprinc}.
As in these related operators, a basic ingredient here will be the
external antisymmetric tensor field
\eq
\tilde{A} _{\mu \nu}(z,x)=(b/4\pi)\int _{T_{x}(S)}d^{3}\xi _{\mu}
\Phi _{\nu}(\xi-x)\delta ^{4}(z-\xi)  -(\mu \leftrightarrow \nu)
\label{ext}
\en
where b is an arbitrary dimensionless real parameter and $\Phi _{\nu}=
(0,0,0,\frac{1-\cos\theta}{r\sin\theta})$ (in the coordinate system
$(t,r,\theta,\varphi)$). The integral in (~\ref{ext}) is performed over
the three-dimensional hypersurface $T_{x}(S)$ whose surface element
$d^{3}\xi _{\mu}$ has only the 0-component nonvanishing. $T_{x}(S)$ is the
region of the {\myfont R}$^{3}$ space external to the surface S at
$z^{0}=x^{0}$. This
surface consists of a piece of sphere  of radius $\rho$
centered at $\vec{x}$ ($0\leq \theta
\leq \pi -\delta$) superimposed to an infinite trunk of
 cone with vertex at $\vec{x}$ and angle $\delta$
(cut a distance $\rho \cos \delta$ from the tip)
 with axis along $\theta =\pi $.

\spa The $\mu -$operator is then constructed in the following way
\eq
\mu (x)=\lim_{\rho ,\delta \rightarrow 0}
\exp \{-(i/6)\int d^{4}z W^{\mu \nu \alpha}
(-\Box)^{-1}\tilde{A}_{\mu \nu \alpha}(z;x)\}
\label{mu}
\en
where $\tilde{A}_{\mu \nu \alpha}$ is the field intensity tensor of
$\tilde{A}_{\mu \nu}$. The operator $\mu $, as it stands, depends on the
hypersurface $T_{x}(S)$. Nevertheless, as it happens in the case of the above
mentioned related operators \cite{kinks,emprinc} all the hypersurface
dependence of the $\mu $ correlation functions can be eliminated
by the introduction of a  renormalization counterterm whose
form is uniquely determined solely by the requirement of
hypersurface invariance.
The parameters $\rho$ and $\delta$ will be used as regulators which
will be eliminated at the end of the calculations.

\spa Inserting (~\ref{ext}) in (~\ref{mu}) and observing that the
surface element in (~\ref{ext}) only possesses the 0-component
nonvanishing we get
\eq
\mu(x) =\lim_{\rho ,\delta \rightarrow 0}\exp \{(ib/
4\pi)\int _{T_{x}(S)}d^{3}
 \vec{\xi} \ W^{0ij}(x^{0},\vec{\xi})
(-\Box)^{-1}\partial _{i}\Phi _{j}( \vec{\xi}- \vec{x})\}
\label{muw}
\en
Using the above solution for W in terms of $A_{\mu}$ we could
also express $\mu$ in terms of the electromagnetic field.
 Observe that inserting
the identity  (which is valid in $T_{x}(S)$)
$
\vec{\nabla}\times \vec{\Phi}\equiv\vec{\nabla}[-\frac{\scriptsize 1}
{\scriptsize |\vec{x}|}]$
in (~\ref{muw}) and thereby expressing $\mu$
in terms of the field configuration of a point charge,
 the singularity along the $\theta =
\pi$ axis would disappear and we could already take the limit $\delta
\rightarrow 0$ safely, thereby eliminating the piece of cone and just
keeping the sphere of radius $\rho$ in the definition of the
surface S.

\spa Let us show now that $\mu $ is indeed a charge (or topological
charge) bearing operator. The charge and topological charge densities,
which are identified by the first equation
 in  (~\ref{ecampo1}) are given, respectively
by $j^{0}=\partial _{i}E^{i}$ and $J^{0}=\epsilon ^{ijk}\partial
 _{i}W_{jk}$. Using the expansion for $e^{A}Be^{-A}$ and the above
 commutation rules,
we find the equal times commutator
$$
[\rho(y),\mu (x)]=(b/4\pi)\mu (x)\lim_{\rho ,\delta\rightarrow 0}
\int_{T_{x}(S)}d^{3}\vec{\xi}\ \epsilon^{ijk}\partial^{(y)}_{k}
\delta^{3}(\vec{\xi}-\vec{y})\partial^{(\xi)}_{i}\Phi_{j}(\vec{\xi}-\vec{x})
$$
where $\rho (x)$ stands either for the charge or topological charge density.
Using the Gauss theorem and the identity involving $\vec{\nabla}\times
\vec{\Phi}$ we can straightforwardly evaluate the above integral (we
can see that the cone piece of S does not contribute) obtaining
$
[\rho (y),\mu (x)]=b\mu (x)
$
This relation shows that $\mu$ does
indeed create states bearing b units of charge.
 A very important relation involving $\mu (x)$ that can be
derived along the same lines
is the commutator with the electric field
$$
[E^{k}(y),\mu (x)]_{ET}=b\frac{(\vec{y}-\vec{x})^{k}}{4\pi |\vec{y}
-\vec{x}|^{3}}\ \  \mu (x)
$$
This implies that the vacuum expectation value of the electric field in
the
 states created by $\mu$ is the
field configuration of a point electric charge of magnitude b:
$$
<\mu (x)|E^{k}(y) |\mu (x)>_{ET}=b\frac{(\vec{y}-\vec{x})^{k}}{4\pi |\vec{y}
-\vec{x}|^{3}}
$$
This expression characterizes $|\mu (x)>$ as a coherent state
of photons.

\spa Let us study now the $\mu$ correlation function. Taking the
expression of $\mu$, Eq. (~\ref{mu}), the lagrangian ${\cal L}$  and going to
euclidean space (in which we will work henceforth) we may write
$$
<\mu (x)\mu ^{\dagger}(y)>=Z^{-1}
\int DW_{\mu \nu}DA_{\mu} \exp \{-\int d^{4}z [{\cal L}+
{\cal L}_{GF}+{\cal L}_{CT}
$$

\eq
-(1/6)W^{\mu \nu \alpha}(-\Box)^{-1}
\tilde{A}_{\mu \nu \alpha}(z;x,y) ]\}
\label{cf1}
\en
where ${\cal L}_{GF}={\cal L}_{GFW}+{\cal L}_{GFA}$ is the gauge
fixing term, with
$${\cal L}_{GFW}=-(\xi _{1}/8)W_{\mu \nu}K^{\mu \nu \alpha \beta}
(-\Box)^{-1}W_{\alpha \beta}$$
where $K^{\mu \nu \alpha \beta}=\partial^{\mu}
\partial^{\alpha}\delta^{\nu \beta}+\partial^{\nu}\partial^{\beta}
\delta^{\mu \alpha}-(\alpha \leftrightarrow \beta )$. The $A_{\mu}$
gauge fixing term is
${\cal L}_{GFA}=-(\xi _{2}/4)A_{\mu}\partial^{\mu}\partial^{\nu}A_{\nu}$.
In (~\ref{cf1}) $\tilde{A}_{\mu \nu \alpha}
(x,y)=\tilde{A}_{\mu \nu \alpha}(x)-\tilde{A}_{\mu \nu \alpha}(y)$
 and ${\cal L}_{CT}$ is the above
mentioned  hypersurface renormalization counterterm to be determined
below.
We see that $<\mu\mu^{\dagger}>=e^{F[\tilde{A}_{\mu \nu}]}$ is the vacuum
functional in the presence of the external field $\tilde{A}_{\mu \nu}$.
This property of the correlation functions of $\mu$
 is common to all the above mentioned topological charge bearing
related operators
\cite{kinks} and
follows from the general fact that  topological charge carrying
operators are closely related to the
 disorder variables of Statistical Mechanics \cite{soldis}. Indeed,
treating these operators as disorder variables \cite{kinks,emprinc}
immediately leads in general \cite{evora}
to a form of $\mu$  which is expressed in terms of
the coupling of the lagrangian field to an external field like
$\tilde{A}_{\mu \nu }$  and also
to a renormalization counterterm consisting in the self-coupling of
this external field
both with the same form as the kinetic term.
 Also here, we will see explicitly that the renormalization
counterterm
$
{\cal L}_{CT}=(1/12)\tilde{A}^{\mu \nu \alpha}(-\Box)^{-1}
\tilde{A}_{\mu \nu \alpha}
$
will absorb the external field infinite self-energy and render the
$\mu $ correlation function hypersurface independent.

\spa Integrating over $A_{\mu}$ and then
 over $W_{\mu \nu}$ in (~\ref{cf1}) with the
help of the euclidean propagator of this field, namely
$$
D^{\mu \nu \alpha \beta}(x)=(1/4)\lim_{m \rightarrow 0}
[(-\Box)\Delta ^{\mu \nu
\alpha \beta}+(1-\xi _{1}^{-1})K^{\mu \nu \alpha \beta}]
[-(1/8\pi ^{2})\ln m\gamma |x|]
$$
where $\gamma$ is the Euler constant and m is an infrared regulator$-$
used to define the inverse Fourier transform of 1/$k^{4}-$ we
obtain the following result
$$
<\mu (x) \mu ^{\dagger}(y)>=\lim_{\rho ,\delta ,m,\epsilon
\rightarrow 0}\exp \{
(b^{2}/2)\sum_{i,j=1}^{2}\lambda_{i}\lambda_{j}
\int_{T_{x_{i}}}d^{3}\xi_{\mu}\partial_{\alpha}^{(\xi)}\Phi_{\nu}
(\vec{\xi}-\vec{x_{i}})
$$

$$
\times\int_{T_{x_{j}}}d^{3}\eta_{\gamma}\partial_{\beta}
^{(\eta)}\Phi_{\rho}(\vec{\eta}-\vec{x_{j}})\ \ \epsilon^{\mu \nu \alpha
 \sigma}\epsilon^{\gamma \rho \beta \lambda}
$$

\eq
\times [ -\Box\delta_{\sigma
\lambda}+\partial^{(\xi)}_{\sigma}\partial^{(\xi)}_{\lambda}]
[-(1/8\pi^{2})\ln\ m\gamma [|\xi -\eta|+|\epsilon |]]-S_{CT}\}
\label{cf4}
\en
In this expression, $x_{1}\equiv x$ , $x_{2}\equiv y$ ,$\lambda_{1}
\equiv +1$ and $\lambda_{2} \equiv -1$. We also introduced the
ultraviolet cutoff $|\epsilon |$ in the euclidean $W_{\mu \nu}$
propagator.
Only the first term of the W-propagator  contributed to (~\ref{cf4}).
In particular, all the gauge dependent terms were cancelled. This happens
because of the gauge invariant way in which the external field is
coupled in (~\ref{cf1}).
The first term in (~\ref{cf4}) is nothing but $S_{CT}$ and therefore
is exactly canceled.
Using Gauss theorem and the identity for $\vec{\nabla}\times\vec{\Phi}$
it is not difficult
to see that the crossed terms (with $i\neq j$) vanish in the limit
$\rho \rightarrow 0$. The self-interaction terms ( with $i=j$) on the
other hand, diverge in this limit. We conclude therefore that the
renormalization counterterm $S_{CT}$ contains {\em only the
unphysical  self interaction
terms}.

 \spa Each of the integrals in
(~\ref{cf4})
can be evaluated straightforwardly by the use of the Gauss
theorem and of the identity for $\vec{\nabla}\times\vec{\Phi}$ . The result is
$
\exp [-F(x-y)+F(\epsilon )]$ where $F(x)=(b^{2}/8\pi^{2})\ln
m\gamma |x|$ and we still must make $m,\epsilon\rightarrow 0$.
Note that the $m\gamma$ factors cancel out. In a charge selection rule
violating correlation function (like $<\mu \mu >$, for instance) we
would have the sign of the $F(x-y)$  term reversed and the
$m\gamma$ factors would no longer cancel, actually forcing the
correlation function to vanish in the limit $m\rightarrow 0$ and
thereby enforcing the charge selection rule.
The ultraviolet divergence at $|\epsilon | \rightarrow 0$
can be eliminated by a multiplicative renormalization of the field
$\mu$, namely $
\mu_{R}(x)=\mu (x)|\epsilon |^{-b^{2}/16\pi^{2}}
$.
 Using this we finally get
$
<\mu_{R}(x)\mu^{\dagger}_{R}(y)>=|x-y|^{-b^{2}/8\pi^{2}}
$
This is our final expression for the $\mu$ field
euclidean correlation function. Observe that $|x-y|$ is a distance
in  4-dimensional euclidean space.
An arbitrary 2n-point correlation
function could be obtained in a straightforward manner by just
inserting additional external fields $\tilde{A}_{\mu \nu}$ in
an expression like (~\ref{cf1}). It would be given by a product
of monomials of the type found in $<\mu \mu^{\dagger}>$. The form
of the $\mu$ correlation functions characterizes it inequivocally
as a local operator.

\spa From the long distance behavior $\lim_
{|x-y|\rightarrow\infty}
<\mu_{R} (x)\mu_{R}^{\dagger}(y)>=0$, we can infer that
 $<\mu (x)>=0$ and therefore
that the states  $|\mu (x)>$ , created by $\mu$ are
orthogonal to the vacuum. The power law decay of the  correlation
function on the other hand implies that these states
are massless.

\spa
As far as we can see, the charge of the states created
 by $\mu $ is not quantized, in the same way as in the two-
dimensional analogous system. It would be interesting to investigate
whether some additional coupling $-$ as it happens with the
Sine-Gordon coupling in the case of the scalar field in 1+1 D $-$
would produce a charge quantization for these states.

\spa Including the coupling of the electromagnetic field
$A_{\mu}$ to a matter current $j^{\mu}_{m}$ (in (~\ref{cf1}) ,
for instance) we would no longer be able to obtain an exact
expression for the $\mu$ correlation functions because the
integration over the matter fields could not be done exactly.
Nevertheless, due to the well known infrared asymptotic freedom
of QED we still may conclude that the long distance behavior of
$<\mu \mu^{\dagger}>$ will be the same. The result
that the operator $\mu (x)$ creates massless states orthogonal to the
vacuum, therefore, also holds in the presence of matter fields.

\spa Let us mention
a well known operator which shares some of the properties
of $\mu$ (as the commutator with the charge operator,
for instance)
, namely, $\exp \{-i(b/4\pi)$ $\int^{x}A_{\mu}dx^{\mu}\}$. This operator
is unacceptable because
for it we would have
nonvanishing string dependent crossed ($i\neq j$) interaction terms
 (see the remarks after Eq.(~\ref{cf4})) which would inevitably
lead to a nonlocal correlation function.

\spa It would be extremely interesting to investigate the conditions
under which the states we found here could be experimentally
observed. These should certainly include the presence of high
intensity electromagnetic fields which would be needed to populate
the coherent photon state. An interesting feature of the massless
 charged states is that they will not radiate since a massless
charge cannot be accelerated.

\spa The results we found in this work show that a physical quantity
which is usually associated with matter, namely, charge, can be
generated as an attribute of some
coherent states of the electromagnetic field itself. It is not
inconceivable that other quantities like spin, mass, flavor, color
and so on could be generated as well as properties of some
peculiar states of the gauge fields in general. This would lead to the
outstanding possibility of describing both matter and the
fields which mediate its interactions within the same unified
framework. We hope this work could provide a small contribution
towards this end.

\vskip 10mm

I would like to thank the Physics Department of Princeton University
and especially C.Callan and D.Gross for the kind hospitality. I am
also very grateful to C.Callan for an interesting discussion. Finally,
 I thank M.A.Mart\'{\i}n-Delgado for teaching me LaTex. This work was
supported in part by the Brazilian National Research Council (CNPq).

\end{document}